%% file: main.tex
\documentclass[%
 reprint,
nofootinbib,
 amsmath,amssymb,
 aps,
]{revtex4-2}

\usepackage{graphicx}
\usepackage{dcolumn}
\usepackage{bm}
\usepackage{theorem}
\usepackage{color}
\usepackage{multirow}
\usepackage{amssymb}
\usepackage{subfigure}
\usepackage{mathrsfs}
\usepackage{hyperref}
\newcommand{\Lop}{\hat L}
\newcommand{\mbp}[1]{{\color{magenta}{#1}}} 
\newcommand{\mbpcom}[1]{\footnote{\color{magenta}{#1}}}

\input{preferences}

\begin{document}

\title{Relativistic Gravity-Induced Entanglement via Frame Dragging}
\author{Eyuri Wakakuwa} 
\email{e.wakakuwa@gmail.com}
\affiliation{Department of Mathematical Informatics, Nagoya University, Furo-cho, Chikusa-ku, Nagoya, 464-8601, Japan}
\author{Luciano Petruzziello}
\email{luciano.petruzziello@uni-ulm.de}
\affiliation{Institut f\"ur Theoretische Physik \& IQST, Albert-Einstein-Allee 11, Universit\"at Ulm, 89069 Ulm, Germany}
\author{Trinidad B. Lanta\~{n}o}
\affiliation{Institut f\"ur Theoretische Physik \& IQST, Albert-Einstein-Allee 11, Universit\"at Ulm, 89069 Ulm, Germany}
\author{Susana F. Huelga}
\email{susana.huelga@uni-ulm.de}
\affiliation{Institut f\"ur Theoretische Physik \& IQST, Albert-Einstein-Allee 11, Universit\"at Ulm, 89069 Ulm, Germany}
\author{Martin B. Plenio}
\email{martin.plenio@uni-ulm.de}
\affiliation{Institut f\"ur Theoretische Physik \& IQST, Albert-Einstein-Allee 11, Universit\"at Ulm, 89069 Ulm, Germany}
\date{\today}

\begin{abstract}
Gravity-induced entanglement has been proposed as a method for testing the non-classical
nature of gravity via tabletop experiments. While most existing proposals are restricted
to the Newtonian limit, the frame dragging effect offers access to genuinely
post-Newtonian features of the gravitational interaction and remains comparatively less
explored. Here, we study gravity-induced entanglement generated by frame dragging in an interferometric setting and compute the entanglement phase between the rotational
degrees of freedom of a source mass and the paths of a particle in two complementary ways:
(i)~via Schr\"odinger evolution with a quantized Lense-Thirring Hamiltonian in the
large angular momentum limit, and (ii)~via the on-shell action of linearized quantum
gravity within the stationary phase approximation. Both approaches yield the same
entanglement phase, consistent with the
proper time difference between the interferometer arms. The path integral derivation
further reveals how gravitational retardation modifies the entanglement phase, 
thereby making the local, relativistically causal linearized-gravity description explicit. Under the standard locality/mediator assumptions used in existing arguments, the resulting entanglement would witness non-classicality of the gravitational interaction. 
\end{abstract}

\maketitle

\section{Introduction}
The opinion that gravity must be described by a quantum theory is far from unanimous; even today, many physicists still question whether there exists a compelling motivation for quantizing the gravitational field beyond the fact that all other interactions are quantum. This debate stretches back to the late 1920s and early 1930s, with the earliest and most systematic attempts carried out by L\'{e}on Rosenfeld \cite{rosenfeld1930gravitationswirkungen, rosenfeld1930quanten}. Rosenfeld pioneered the canonical quantization of the linearized Einstein equations, laying the groundwork for what would become the quantum-gravity program.

By 1950, however, Rosenfeld himself had become increasingly skeptical of the necessity of a quantum theory of gravity. This skepticism became explicit at the 1957 Chapel Hill Conference \cite{Dewitt2011TheRO}, one of the most influential meetings on gravitation of the 20th century. During the discussion sessions, not only Rosenfeld expressed doubts about the necessity of a quantization scheme for gravity, but also others, who stressed that the motivation for quantizing gravity lacked the empirical force that had driven the quantization of electromagnetism.
For Richard Feynman, the fact that gravity must be quantized was natural, which he illustrated by a simple gedanken experiment: Consider a ball heavy enough to source a sizable gravitational field and put it in quantum superposition of two locations; how would a test mass move in response? While Feynman insisted that such a situation requires a fully quantum treatment, many of his colleagues were not entirely convinced. Bondi, for example, was concerned that gravitostatics should not be quantized, and only radiative degrees of freedom should, in an analogy to quantum electrodynamics. 


The healthy skepticism surrounding Feynman's gedanken experiment shows that the quantization of gravity is a subtle question, one that cannot be settled by analogy alone and must be approached from multiple conceptual and experimental angles. 
A well-known example of this subtlety is the Colella-Overhauser-Werner (COW) experiment \cite{colella1975observation}, where neutrons acquire a gravity-induced phase shift fully explained by the Schr\"{o}dinger equation with a classical gravitational potential. Likewise, classical magnetic fields produce Larmor precession of spins without requiring the field itself to be quantized. Such cases highlight that quantum interference produced by a gravitational field does not, by itself, imply quantization of the field, a crucial distinction when interpreting proposed quantum-gravity signatures.

In recent years, the quantum information community has identified a sharper criterion to probe the quantumness of gravity. The key insight is that the gravitational interaction can be viewed as a physical channel through which information is exchanged and correlations are established. If this channel could be simulated using only local operations (LO) on each subsystem, supplemented by the exchange of classical communication (CC), it would belong to the class of LOCC channels. Crucially, LOCC channels cannot generate entanglement. Therefore, if two massive systems initially prepared in a separable state, become entangled solely through their mutual gravitational interaction, the effective gravitational channel cannot be described by LOCC alone~\cite{Lami2024,KafriTaylor2013,KafriTaylorMilburn2014}. In this operational sense, entanglement generation witnesses a non-classical feature of the gravitational interaction as a channel~\cite{Lindner2005,bose2017spin,MarlettoVedral2017,Lami2024}. 

This insight ignited an active research program on gravity-induced entanglement (GIE) \cite{Lindner2005,KafriTaylor2013,KafriTaylorMilburn2014, bose2017spin, MarlettoVedral2017}, and sparked legitimate and important questions about whether any physical field that acts as a quantum channel must necessarily be described by a full quantum field theory (QFT), and what such an observation would say about the quantum nature of the field mediator. 
These questions have been explored from different perspectives, ranging from information-theoretic and quantum field-theoretic approaches to critical analyses of the underlying assumptions~\cite{MarlettoVedral2018,MarlettoVedral2020,HallReginatto2018,Marshman2020,Galley2022,Belenchia2018,Rydving2021,Carney2022,Danielson2022,MartinMartinezPerche2023,christodoulou2019possibility,christodoulou2023gravity,ChristodoulouRovelli2020,Chen2023quantumstatesof,Fragkos2022,AzizHowl2025,DiBiagio2025ClassicalEntangle,MarlettoVedral2025ClassicalGravity,mitrakos2026does,trillo2025diosi,spaventa2023tests,anastopoulos2021gravitational,ludescher2026gravity,weller2025there,angeli2025entanglement,oppenheim2023covariant,marletto2025classical,aspelmeyer2026quantum,feng2026collapse,gundhi2026can,schneider2025demonstration,lin2026can,diosi2025no,boulle2025subsystems,xue2026aziz, hall2018two}. For broader reviews on GIE, see Refs.~\cite{Huggett2023,BoseReview2025,CarneyStampTaylor2019}. 

Most existing proposals consider two particles, each initially prepared in a spatial superposition $(\ket{L}_A+\ket{R}_A)\otimes(\ket{L}_B+\ket{R}_B)$, where $\ket{L}_{A(B)}$ and $\ket{R}_{A(B)}$ denote two possible localized positions of particle $A(B)$. The system then evolves under a first-quantized Newtonian interaction of the form $\hat U_G=\exp\left[-({i}/{\hbar})\int dt\,{Gm_A m_B}/{|\hat x_A-\hat x_B|}\right]
$, where $m_{A(B)}$ and $\hat{x}_{A(B)}$ are the masses and positions of particles A(B).
Since each component of the joint superposition corresponds to a different relative distance between the particles, the Newtonian interaction assigns a different phase to each configuration, \emph{i.e.}, to each branch of the joint spatial state. These relative phases can entangle the spatial degrees of freedom of the two particles. 

While this non-relativistic description is natural from a quantum information theory perspective, it is more problematic from the viewpoint of high-energy physics, where one would usually expect interactions to be mediated by local field degrees of freedom. This is a reasonable concern, as even Feynman had to address this point at the Chapel Hill conference: ``\textit{If I write down a term ${Gmm'}/r_{ij}$ in the equation for the universe and if I have only particles in the system with no question of field variable, is this a quantum theory of gravitation or is it a classical theory? That’s a question of definition. In other words, when this term is included has the gravitation field been quantized?}"~\cite{Dewitt2011TheRO}.

Aiming to clarify the assumptions under which the Newtonian GIE description is derived from linearized quantum gravity, Christodoulou \emph{et al.}~\cite{christodoulou2023locally} showed that the same relative phases obtained from the first-quantized Newtonian interaction can be derived from the on-shell action of linearized quantum gravity in the path-integral formalism. In this formulation, spacetime locality is kept manifest, and retardation effects in the entanglement generation process are incorporated. The Newtonian result emerges only after taking the near-field and slow-motion limits. It was pointed out in \cite{mitrakos2026does} that observing such retardation effects would strengthen the inference from the detection of GIE for the existence of gravitons.

However, this does not by itself make the interpretation model-independent. As emphasized by Fragkos, Kopp, and Pikovski~\cite{Fragkos2022}, the same entangling dynamics can in principle be described either through quantum mediating field degrees of freedom or through non-local interaction processes. Thus, observing entanglement would rule out a purely classical LOCC description of the interaction, but the stronger claim that it directly witnesses a quantized gravitational mediator relies on the additional assumption of locality.

The observation of a Newtonian GIE phase would undoubtedly be an exciting result, shedding light on quantum aspects of gravity in a concrete experimental setting. Nevertheless, since this phase survives the formal limit $c\to\infty$, the conclusions that can be drawn from it are confined to the leading non-relativistic, near-field regime of the gravitational interaction. A natural next step is therefore to retain finite-$c$ contributions in the weak-field expansion of general relativity.

Some of the authors of the present work have proposed such a step by considering entanglement generation through the frame dragging contribution of linearized general relativity~\cite{lantano2024low,Petruzziello2025}. This interaction is associated with the gravitomagnetic sector of the weak gravitational field and is sourced by angular momentum rather than mass density alone. In this setting, the effective interaction takes the form of a dipole-dipole-like coupling between the angular momenta of two rotating microspheres. Such an experiment would probe quantum aspects of a genuinely general relativistic interaction.

At the same time, this should not be overstated. The proposed effect remains a near-field weak-gravity interaction, and observing the corresponding entanglement would not by itself amount to a direct detection of propagating gravitational degrees of freedom. Rather, it would extend GIE tests from the Newtonian sector to a finite-$c$, gravitomagnetic sector of linearized gravity. The inference that the entanglement was mediated by local quantized gravitational degrees of freedom would therefore still require the additional locality assumption discussed above.
In a related direction, Ref.~\cite{wakakuwa2025detectability} studied gravity-induced entanglement in quantum clock interferometry \cite{zych2011quantum,roura2020gravitational}, where the frame-dragging contribution appears as a post-Newtonian correction to the proper-time difference accumulated along the two interferometer arms.
From a different perspective, Ref.~\cite{chen2024quantum} showed, within the framework of linearized quantum gravity, that interactions between delocalized quantum sources of gravity can give rise to entanglement that cannot be accounted for by the Newtonian potential alone.

Historically, one can roughly identify three stages in the research of GIE: (1)  Effective entangling interactions: The original proposals \cite{Dewitt2011TheRO, Lindner2005,KafriTaylor2013} and the more recent Bose-Marletto-Vedral \cite{bose2017spin, MarlettoVedral2017} works established that
gravitational interactions could generate entanglement. At this level, gravity essentially enters
through an effective interaction Hamiltonian and the focus is on entanglement generation itself. (2) Local mediator interpretation: Christodoulou \textit{et al.} \cite{christodoulou2023locally} demonstrated that the familiar Newtonian entangling phase admits a derivation from a local and relativistically causal linearized gravity description. This was conceptually important because it showed that the Newtonian result was not merely an artefact of an instantaneous-looking potential but could be embedded within a local field-theoretic
framework. In other words, the Newtonian result was shown to be consistent with a local mediator picture. (3) Beyond the Newtonian regime: This is where the present work belongs. We demonstrate that the correspondence established by Christodoulou \textit{et al.} \cite{christodoulou2023locally} is not merely
a special property of the Newtonian regime. It survives when moving to the post-Newtonian
frame dragging contribution, where the interaction is sourced by angular momentum and
encoded in the gravitomagnetic sector of the weak gravitational field. 

More precisely, we recover the same entangling phase from two complementary descriptions: the Schr\"{o}dinger evolution generated by a quantized Lense-Thirring Hamiltonian, and the on-shell action of linearized quantum gravity coupled to matter. This agreement shows that the effective quantum description is consistent with a local, relativistically causal field-theoretic treatment of linearized gravity. Viewed from this perspective,  the central question addressed by this work is not whether
one can we calculate the phase in two different ways but rather: \textit{``Does the correspondence between effective gravitational entangling Hamiltonians and local
linearized gravity descriptions persist beyond the Newtonian regime?"}.

Furthermore, the path integral formalism used here makes the
causal structure explicit through the retarded and advanced switching functions. In this sense, the path integral derivation seems to suggest a hierarchy of increasingly strong tests: (1) Observation of entanglement $\rightarrow$ (2) Observation of the predicted phase $\rightarrow$ (3) Observation of retardation-induced corrections $\rightarrow$ (4) Verification that these corrections follow the predictions of a local relativistic mediator.

The paper is structured as follows. In Sec.~\ref{EntSch}, we re-derive the entanglement phase generated by frame dragging within a non-relativistic framework by simply promoting the Lense-Thirring Hamiltonian to a quantum operator acting on the angular momentum eigenstates. In Sec.~\ref{sec:PathInt}, we obtain the same entangling phase from the on-shell action of linearized quantum gravity and discuss how relativistic retardation manifests in the amount of generated entanglement. The detailed calculations of the relevant action integrals are collected in Appendix~\ref{app:calaction}. Finally, Sec.~\ref{sec:concl} contains discussions and conclusions.

{\it Notations:}
Throughout this paper, the speed of light is denoted by $c$, the gravitational constant by $G$, and the reduced Planck constant by $\hbar$.
We use the signature $(-+++)$ for the spacetime metric $g_{\mu\nu}$. 
Greek indices $\{\mu,\nu\}$ run over $0,1,2,3$, while Latin indices $\{i,j\}$ run over spatial components $1,2,3$.
We also adopt the Einstein summation convention, e.g., $p_\mu q^\mu \equiv \sum_{\mu=0,1,2,3} p_\mu q^\mu$.

\section{Entanglement from the Schr\"{o}dinger equation}
\label{EntSch}

The setup we consider here closely parallels the one analyzed in Ref.~\cite{wakakuwa2025detectability} by one of the authors.
Here, a quantum clock interferometer \cite{zych2011quantum,roura2020gravitational} is employed to probe post-Newtonian gravity in the field of a rotating source mass, in particular the proper-time difference induced by the gravitomagnetic clock effect \cite{cohen1993standard,ciufolini1995gravitation}. 
In that work, by virtue of the symmetry of the configuration, the leading Newtonian contribution cancels between the two arms of the interferometer, leaving the frame-dragging effect as the dominant signature. In so doing, both a single-particle interferometric scheme to detect the rotating gravitational field and a gravity-induced entanglement scheme between source and clock are discussed. 

Here, we focus more in detail on the latter aspect of such a configuration and compute the resulting entangling phase in two complementary ways.
Unlike in \cite{wakakuwa2025detectability}, we do not incorporate the internal degree of freedom of the particle and instead consider an atom interferometric setup rather than quantum clock interferometry.
This simplification allows for a more explicit characterization of the entanglement in terms of the entangling phase.

To this aim, we first describe the above setting in a non-relativistic framework by promoting the Lense-Thirring Hamiltonian to a quantum operator. We consider the usual eigenstates of angular momentum $\ket{l, m}$, which satisfy $\hat{L}_z \ket{ l, m} =  m  \ket{l,m}$ and $\hat{\vec{L}}^2 \ket{ l, m}=l( l+1)\ket{l, m}$. Here, we made use of dimensionless angular momentum operators, as the right dimensions will be introduced in the Hamiltonian directly. This means that the angular momentum $l$ and its quantum number $\tilde l$ are related via the simple rescaling $l=\hbar \,\tilde l$.

\begin{figure}[t]
\centering
\includegraphics[bb={0 0 315 301}, scale=0.7]{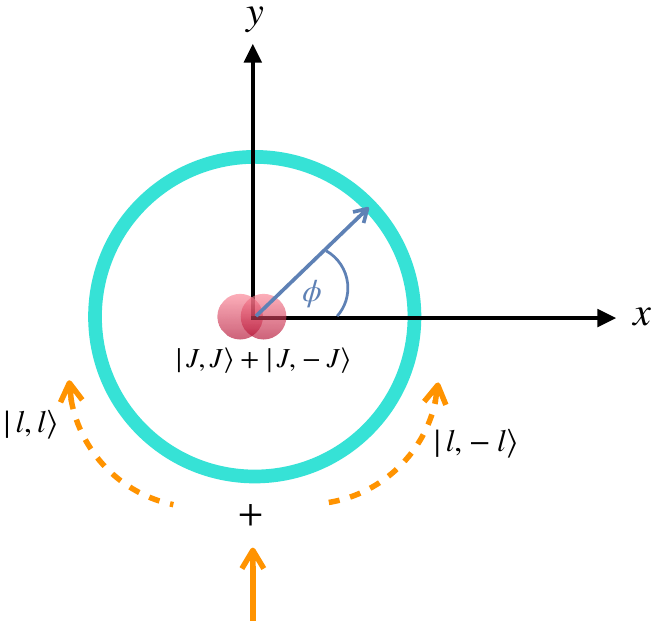}
\caption{Schematic picture of the considered setting, involving a source at the center of the circular interferometer in a superposition of different angular momentum states along the $z$ axis and a probe particle in a spatial superposition along the arms of the interferometer. The said spatial superposition is maximally entangled with the orbital angular momentum of the probe particle along the $z$ axis, which thus is itself in a superposition.}\label{fig:superposition_clockinterf}
\end{figure}

Furthermore, we assume that, before the interaction, the state of the source and probe 
factorizes as
\begin{align}\label{eq:instate}
\ket{\psi (t=0)} = \frac{1}{2}\left( \ket{\tilde J,\tilde J}_S + \ket{\tilde J,-\tilde J}_S\right)\otimes \left(\ket{\tilde l,\tilde l}_P + \ket{\tilde l,-\tilde l}_P \right)
\end{align}
where the kets corresponding to the source (probe) carry the subscript S(P). Note that we describe the probe with the $\mathrm{SU}(2)$ states $\ket{\tilde l,\pm\tilde l}_P$; we show below that, in the large angular momentum limit, the full dynamics is confined to the two-dimensional subspace spanned by the extremal states of both the source and the probe. 

Now, we assume the source to be prepared in a superposition of two angular momentum states along $z$, up and down, with amplitude $J$. Analogously, the probe particle is sent in a circular interferometer, where a beam splitter sends the particle either to the right or to the left, giving it an angular momentum of $l$ or $-l$ along $z$ (see Fig~\ref{fig:superposition_clockinterf}). Since both particles have angular momentum, they source a gravitational field which enters the total Hamiltonian as~\cite{lantano2024low,Petruzziello2025}
\begin{align}\label{eq:Ht}
    \hat{H} & =\, \frac{\hbar^2\hat{L}^2_{S}}{2I_S} + \frac{\hbar^2\hat{L}^2_{P}}{2I_P} - \frac{G M m_0}{\tilde{d}}
    \nn\\&\quad-\frac{2G \hbar^2}{c^2 \tilde{d}^3} \left[\hat{\vec{L}}_S\cdot \hat{\vec{L}}_P - 3(\hat{\vec{L}}_S \cdot \vec{r})(\hat{\vec{L}}_P \cdot \vec{r}) \right], 
\end{align}
where $I_S$ and $I_P$ are the moments of inertia, $\tilde{d} := {r + |\hat{\vec{r}}_S- \hat{\vec{r}}_P}| $ with $r$ the distance between the centers of mass and $\hat{\vec{r}}_i$ the displacement of the $i$-th system from its equilibrium position. The dimensionless $\vec{L}_P$ and $\vec{L}_S$ (we already factored out an $\hbar$) denote the angular momentum of the probe with mass $m_0$ and the source with mass $M$, respectively\footnote{The additional factor $2$ appearing in the expression is related to the fact that, differently from Refs.~\cite{wald,gravitation}, the orbital angular momentum is considered rather than the spin.}, and 
$\vec{r}$ is the unit vector connecting the two objects \cite{wald,barker1975gravitational,gravitation}. While other relativistic corrections
exist in a fully post-Newtonian expansion of the gravitational potential \cite{weinberg}, they can be neglected when focusing on entanglement between rotational degrees of freedom. Note that, when considering two extended objects, a multipole expansion of the Newtonian potential is in order. {The impact of} deviations from spherical symmetry is studied in Ref. \cite{lantano2024low} by considering the leading-order contributions.

Since $\hat{L}^2$ commutes with all its components, in the interaction picture with respect to the free (kinetic) part the Hamiltonian, Eq. \eqref{eq:Ht} reads
\begin{align}\label{eq:HI}
\hat{H}_{I} (\vec{r}_\pm)= - \frac{2G \hbar^2}{c^2 R^3}  \left(\hat{\vec{L}}_S\cdot \hat{\vec{L}}_P - 3(\hat{\vec{L}}_S \cdot \vec{r}_\pm)(\hat{\vec{L}}_P \cdot \vec{r}_\pm)\right), 
\end{align}
where we made the dependence on the connecting vector $\vec{r}_\pm$ explicit, with the subscript denoting the path, \textit{i.e.} whether the angular momentum along $z$ is $+ l$ or $-l$, and since the source is in the center of the circular path, the distance between the source and each branch of the probe is the same, which we denote as $R$. Depending on the two branches of the probe we have $ \vec{r}_\pm = \pm \cos{\phi}\,\vec{e}_x + \sin{\phi}\,\vec{e}_y $, where $\phi$ is the angle between $\vec{r}_\pm$ and the $x$ axis. Note that, while the probe is in a spatial superposition (which is maximally entangled with its orbital angular momentum), the source is not. We only focus on entanglement between rotational degrees of freedom. With this, Eq. \eqref{eq:HI} then becomes
\begin{align}
 &\hat{H}_{I} (\vec{r_\pm}) \nn\\
 &= - \frac{2G \hbar^2}{c^2 R^3} \left[ \Lop_{Sx} \Lop_{Px} + \Lop_{Sy} \Lop_{Py} + \Lop_{Sz} \Lop_{Pz}\right. \nn\\
 &\quad\left.- 3 \left(\pm \Lop_{Sx} \cos{\phi} + \Lop_{Sy}\sin{\phi} \right) \left(\pm \Lop_{Px} \cos{\phi} + \Lop_{Py}\sin{\phi} \right)  \right] \nonumber \\[2mm]  &= \frac{2G \hbar^2}{4 c^2 R^3}  \left( \Lop_{S+} \Lop_{P-} + \Lop_{S-}\Lop_{P+}  - 4\Lop_{Sz}\Lop_{Pz} \right.\nn\\
 &\quad\left.+ 3 \Lop_{S+}\Lop_{P+} \text{e}^{\mp  2 i \phi} + 3 \Lop_{S-}\Lop_{P-} \text{e}^{\pm  2 i \phi}  \right)\, ,\label{eq:HIfin}
\end{align}
where in the last equality we use the ladder operators defined as $\hat{L}_{x} = (\hat{L}_+ + \hat{L}_-)/2$, $\hat{L}_{y} = (\hat{L}_+ - \hat{L}_-)/(2i)$. 

Now, since the action of the dimensionless $\hat L_{\pm}$ on a generic angular momentum state is 
\begin{equation}
    \hat L_\pm|\tilde l,\tilde m\rangle=\sqrt{(\tilde l\mp\tilde m)(\tilde l\pm \tilde m+1)}|\tilde l,\tilde m\pm1\rangle,
\end{equation}
one can consider the regime where $\sqrt{\tilde J \tilde l} \gg 1 $ (\emph{i.e.}, the regime with a high number of angular momentum quanta). It turns out that the above approximation greatly simplifies the action of the operator~\eqref{eq:HIfin} on the initial state~\eqref{eq:instate}. Indeed, by virtue of this assumption, the only non-negligible contribution that goes like $\mathcal{O}(\tilde J \tilde l)$ is provided by the action of the $\hat L_{Sz}\hat L_{Pz}$ term in $\hat H_I$ (as also argued in Ref.~\cite{lantano2024low}). Hence, the final state differs from~\eqref{eq:instate} only because of a relative phase, that is
\begin{align}
    |\psi(t)\rangle
    &=\frac{1}{2}\left[|\tilde J,\tilde J\rangle_S|\tilde l,\tilde l\rangle_P+|\tilde J,-\tilde J\rangle_S|\tilde l,-\tilde l\rangle_P\right.\nn\\
    &\quad\left.+e^{-4{i\frac{G\hbar t\tilde J\tilde l}{R^3c^2}}}\left(|\tilde J,\tilde J\rangle_S|\tilde l,-\tilde l\rangle_P+|\tilde J,-\tilde J\rangle_S|\tilde l,\tilde l\rangle_P\right)\right].
    \label{eq:psitLT}
\end{align}
The emergence of this relative phase generates non-vanishing quantum correlations between the angular momentum of the source and the probe. The entanglement phase $\Delta $ can then be rewritten in terms of the dimensionful quantities as
\begin{equation}\label{eq:entph}
    \Delta=\frac{4GJlt}{\hbar R^3 c^2}\,.
\end{equation}
However, we know that $l=Rm_0v$, with $v$ being the probe's velocity. In addition to that, since the motion is assumed to be uniformly circular, we can write $v=\omega R=2\pi \nu R$, with $\omega$ being the angular velocity and $\nu$ the frequency. Now, as the dynamics starts and ends in two points separated by a diameter $2R$ (see Fig.~\ref{fig:superposition_clockinterf}), the total time $t$ is half the period, meaning that Eq.~\eqref{eq:entph} can be cast in the final form
\begin{equation}\label{eq:entph2}
    \Delta=\frac{4\pi GJm_0}{\hbar R c^2}\,.
\end{equation}
At this point, the above expression must be compared to the one that can be derived from a full-fledged path integral calculation.


\section{Entanglement from linearized quantum gravity and path integral}
\lsec{PathInt}

\subsection{Energy-momentum tensor and metric perturbation}

In the linearized approximation of Einstein's gravity, the metric perturbation $h_{\mu\nu}$ is obtained from the energy-momentum tensor $T_{\mu\nu}$ as a solution of the following wave equation: $\Box h_{\mu\nu}=-{16\pi G}\bar{T}_{\mu\nu}/c^4$,
where $\bar{X}_{\mu\nu}=X_{\mu\nu}-\eta_{\mu\nu}(\eta^{\rho\sigma}X_{\rho\sigma})/2$, and the metric is given by $g_{\mu\nu}=\eta_{\mu\nu}+h_{\mu\nu}$.
The retarded solution of this equation is given by
\alg{
h_{\mu\nu}(t,{\bm r})=\frac{4G}{c^4}\int d^3{\bm r}'\frac{\bar{T}_{\mu\nu}({\bm r}',\tilde{t})}{|{\bm r}-{\bm r}'|},
\laeq{retardedh}
}
where $\tilde{t}\equiv\tilde{t}(t,{\bm r},{\bm r}')$ is the retarded time defined by $c\tilde{t}=ct-|{\bm r}'-{\bm r}|$.
In our case, the energy-momentum tensor can be written as the sum of two terms, one caused by the source mass and the other due to the probe particle:
$T_{\mu\nu}=T_{\mu\nu}^{\rm source}+T_{\mu\nu}^{\rm probe}$.
Likewise, the metric perturbation is separated into $h_{\mu\nu}=h_{\mu\nu}^{\rm source}+h_{\mu\nu}^{\rm probe}$.

The trajectory of the probe particle is parametrized by the coordinate time $t$ as ${\bm q}(t)\equiv(x_p(t),y_p(t),z_p(t))$, and the velocity by ${\bm v}(t)\equiv(v_x(t),v_y(t),v_z(t))$. 
Since the trajectory is confined in the $xy$-plane, it is assumed that $z_p(t)=0$ and $v_z(t)=0$.
The Lorentz factor is
$\gamma(t)=1/\sqrt{1-v(t)^2/c^2}$, where $v(t)^2=v_x(t)^2+v_y(t)^2+v_z(t)^2$.
The rest mass of the particle is denoted by $m_0$.
The energy-momentum tensor of the probe particle can be represented as
\alg{
T_{\rm probe}^{\mu\nu}(t,{\bm r})={m_0}\delta^{(3)}[{\bm r}-{\bm q}(t)]V_{\rm probe}^{\mu\nu}(t),
\laeq{Tclock}
}
where $V_{\rm probe}^{\mu\nu}(t)=\gamma(t)v^\mu(t)v^\nu(t)$ and $(v^\mu(t))\equiv(c,v_x(t),v_y(t),v_z(t))$.
From this, the metric perturbation due to the probe particle is given by
\alg{
h_{\mu\nu}^{\rm probe}(t,{\bm r})
=\frac{4Gm_0}{c^4}\int d^3{\bm r}'\frac{\delta^{(3)}[{\bm r}'-{\bm q}(\tilde{t})]\bar{V}_{\mu\nu}^{\rm probe}(\tilde{t})}{|{\bm r}-{\bm r}'|},
}
where $\tilde{t}\equiv\tilde{t}(t,{\bm r})$ is the solution of $c(t-\tilde{t})=|{\bm r}-{\bm q}(\tilde{t})|$.
Following the lines of \cite{christodoulou2023locally} (see Appendix therein), it can be shown that
\alg{
h_{\mu\nu}^{\rm probe}(t,{\bm r})
&=\frac{4Gm_0}{c^4}\frac{\bar{V}_{\mu\nu}^{\rm probe}(\tilde{t})}{|\tilde{{\bm d}}|-\tilde{{\bm d}}\cdot{\bm v}(\tilde{t})/c},
\laeq{hclock}
}
where $\tilde{{\bm d}}=\tilde{{\bm d}}(t,{\bm r})={\bm r}-{\bm q}(\tilde{t}(t,{\bm r}))$.

For the energy-momentum tensor of the source mass, we consider
\alg{
&T_{00}^{\rm source}=c^2M\delta^{(3)}({\bm r}),
\quad
T_{ij}^{\rm source}=0,
\quad
T_{30}^{\rm source}=0
\laeq{sourceT1}
}
and
\alg{
 T_{10}^{\rm source}&=-\frac{1}{2}cJ(t)\partial_y\delta^{(3)}({\bm r}),\\
T_{20}^{\rm source}&=\frac{1}{2}cJ(t)\partial_x\delta^{(3)}({\bm r}),
\laeq{sourceT2}
}
as in Ref.~\cite{gullu2014spin}.
Here, we incorporated the time dependence of the angular momentum of the source mass as $J(t)\equiv J\Theta(t)$, where the function $\Theta(t)$ describes switch-on/off of the superposition of the source's angular momentum.
A simple calculation yields 
$
\bar{T}_{\mu\mu}^{\rm source}=c^2M\delta^{(3)}({\bm r})/2\:(\mu=0,1,2,3)
$
and $\bar{T}_{\mu\nu}^{\rm source}=T_{\mu\nu}^{\rm source}\:(\mu\neq\nu)$.
Substituting this to Eq.~\req{retardedh}, we obtain
\alg{
&h_{\mu\mu}^{\rm source}=\frac{2GM}{c^2|{\bm r}|}\quad(\mu=0,1,2,3),\laeq{hsource1}\\
&h_{01}^{\rm source}=\frac{2GJ\Theta(t-|{\bm r}|/c)y}{c^3|{\bm r}|^3},
\\
&h_{02}^{\rm source}=-\frac{2GJ\Theta(t-|{\bm r}|/c)x}{c^3|{\bm r}|^3},\laeq{hsource2}\\
&h_{03}^{\rm source}=h_{ij}^{\rm source}=0\;(i\neq j).\laeq{hsource3}
}
The switching of the source's angular momentum, described by the function $\Theta(t)$, may in general be accompanied by gravitational radiation.
The emitted gravitons generate a gravitational field that carries which-branch information of the source's angular momentum, leading to decoherence of its superposition.
To avoid a complex analysis of this effect, we assume that the switching is done adiabatically, so that the radiation is negligible.
Mathematically, this corresponds to assuming that $\Theta(t)$ is a slowly varying function.
We also assume that the branch-dependent backaction on the apparatus used to prepare the superposition is negligible, as is commonly assumed in the literature on gravity-induced entanglement.

\subsection{Action integral}

The on-shell action of linearized quantum gravity coupled to matter is given by
\alg{
S=\frac{1}{4}\int dtd^3{\bm r} (h_{\mu\nu}^{\rm source}+h_{\mu\nu}^{\rm probe})(T^{\mu\nu}_{\rm source}+T^{\mu\nu}_{\rm probe}).
}
As assumed in \cite{christodoulou2023locally}, the terms corresponding to the self energy can be dropped. Hence, the total action integral is decomposed as $S=S^{\rm sp}+S^{\rm ps}$,
where
\alg{
S^{\rm sp}&=\frac{1}{4}\int dtd^3{\bm r}\:h_{\mu\nu}^{\rm source}T^{\mu\nu}_{\rm probe},\laeq{actionintsc}\\
S^{\rm ps}&=\frac{1}{4}\int dtd^3{\bm r}\:h_{\mu\nu}^{\rm probe}T^{\mu\nu}_{\rm source}.\laeq{actionint}
}
The component $S^{\rm sp}$ represents the redshift of the probe particle caused by the gravitational field generated by the source mass.
$S^{\rm ps}$ corresponds to the gravitational back-action of the probe particle on the source mass.

Based on the expressions presented above, we prove in \rApp{calaction} that, in the slow-motion approximation in which terms of order $v/c$ and higher are neglected,
 we have $S^{\rm sp}=S_0^{\rm sp}+S_J^{\rm sp}$, where $S_0^{\rm sp}$ is independent of the rotation of the source mass and 
\alg{
S_J^{\rm sp}
&=
-\frac{GJm_0}{c^2}\int 
\Theta(t^{\rm sp}({\bm q}))\frac{({\bm q}\times d{\bm q})_z}{|{\bm q}|^3}.
}
Here, we defined the retarded time $t^{\rm sp}({\bm q})\equiv t({\bm q})-|{\bm q}|/c$, with $t({\bm q})$ being the inverse function of ${\bm q}(t)$, and the subscript $z$ denotes the $z$-component of the vector.
Likewise, we have $S^{\rm ps}=S_0^{\rm ps}+S_J^{\rm ps}$, where
\alg{
S_J^{\rm ps}
=
-\frac{GJm_0}{c^2}\int 
\Theta(t^{\rm ps}({\bm q}))\frac{({\bm q}\times d{\bm q})_z}{|{\bm q}|^3}
}
and the advanced time is defined as $t^{\rm ps}({\bm q})\equiv t({\bm q})+|{\bm q}|/c$.
Note that the trajectory of the probe particle is confined in the $xy$ plane and is represented as a function of the longitude $\phi$ as ${\bm q}(\phi)$, with which $t({\bm q}(\phi))$ will henceforth be abbreviated as $t(\phi)$, and $({\bm q}\times d{\bm q})_z=|{\bm q}|^2d\phi$.

In total, the action integral is given by $S_{\rm tot}=S_0+S_J$, where
\alg{
&S_J=S_J^{\rm sp}+S_J^{\rm ps\nonumber}\\
&=-\frac{GJm_0}{c^2}\int 
\left[\Theta\left(t^{\rm sp}(\phi)\right)+\Theta\left(t^{\rm ps}(\phi)\right)\right]\frac{d\phi}{|{\bm q}(\phi)|}.
}
The difference of the action integral between the two paths is, due to the symmetry of the setup, given by
\alg{
&\Delta S
=
S_{-J}-S_{+J}
\nn\\
&
=
\frac{2GJm_0}{c^2}\int 
\left[\Theta\left(t^{\rm sp}(\phi)\right)+\Theta\left(t^{\rm ps}(\phi)\right)\right]\frac{d\phi}{|{\bm q}(\phi)|}.
\laeq{DeltaSsccs}
}
In particular, if the rotation of the source mass is switched on at a sufficiently early time and is switched off sufficiently late, i.e., $\Theta=1$ at any time during the interaction, we have
\alg{
\Delta S
=
\frac{4GJm_0}{c^2}\int 
\frac{d\phi}{|{\bm q}(\phi)|}.
\laeq{DeltaST1}
}
This coincides with the action integral evaluated from the proper-time difference between the paths, based on the result in \cite{wakakuwa2025detectability}.

\subsection{Entanglement phase}
\lsec{TranA}

Let $|\!\uparrow\rangle$ and $|\!\downarrow\rangle$ denote the rotation of the source mass, corresponding to clockwise and counterclockwise, respectively.
Furthermore, let $|L\rangle$ and $|R\rangle$ denote the left and right paths of the probe particle. 
In the same way as in \cite{christodoulou2023locally}, we apply the stationary phase approximation to both the field and path degrees of freedom.
Then, the degree of freedom of the gravitational field will be uncorrelated from the source mass and the probe particle after the whole interaction.
Thus, the joint evolution of the source's rotational degree of freedom, the path and position of the probe particle is described by the unitary operator
\alg{
U=&e^{\frac{iS_{-J}}{\hbar}}(\ket{\uparrow}\!\bra{\uparrow}\otm\proj{L}+\ket{\downarrow}\!\bra{\downarrow}\otm\proj{R})\nn\\
&+e^{\frac{iS_{+J}}{\hbar}}(\ket{\uparrow}\!\bra{\uparrow}\otm\proj{R}+\ket{\downarrow}\!\bra{\downarrow}\otm\proj{L}).
\laeq{unitarywhole}
}
Suppose that the rotation of the source mass is initially prepared in an equal superposition of the two opposite orientations $(|\!\!\uparrow\rangle+|\!\!\downarrow\rangle)/\sqrt{2}$.
The initial state of the whole system is then given by
\alg{
\ket{\Psi_i}=\frac{1}{\sqrt{2}}\left(|\!\uparrow\rangle+|\!\downarrow\rangle\right)\otm\frac{1}{\sqrt{2}}\left(\ket{L}+\ket{R}\right).
}
Applying the unitary evolution \req{unitarywhole} to the above, the state becomes 
\alg{
\ket{\Psi_f}=
&
\frac{1}{2}\left[|\!\uparrow\rangle\left(e^{\frac{iS_{-J}}{\hbar}}\ket{L}+e^{\frac{iS_{+J}}{\hbar}}\ket{R}\right)\right.\nn\\
&\quad\quad\left.+|\!\downarrow\rangle\left(e^{\frac{iS_{+J}}{\hbar}}\ket{L}+e^{\frac{iS_{-J}}{\hbar}}\ket{R}\right)\right]\nonumber\\
=
&
\frac{1}{2}e^{\frac{iS_{+J}}{\hbar}}\left[|\!\uparrow\rangle\left(e^{\frac{i\Delta S}{\hbar}}\ket{L}+\ket{R}\right)\right.
\nn\\
&\quad\quad\quad\quad\left.+|\!\downarrow\rangle\left(\ket{L}+e^{\frac{i\Delta S}{\hbar}}\ket{R}\right)\right].
\laeq{expPsif}
}
Note that the factor $\exp[{iS_{+J}}/\hbar]$ in the front contributes only a global phase and can be ignored.
The entangling phase is thus given by $\Delta=\Delta S/\hbar$.

We adapt the above formula to the scenario presented in Sec. \ref{EntSch}. 
internal ground state.
Noting that $|{\bm q}(\phi)|=R$, it follows from Eq.~\req{DeltaSsccs} that the entanglement phase is obtained as
\alg{
\Delta
=
\frac{2GJm_0}{\hbar c^2R}\int_{-\frac{\pi}{2}}^{\frac{\pi}{2}}  
\left[\Theta\left(t^{\rm sp}(\phi)\right)+\Theta\left(t^{\rm ps}(\phi)\right)\right]d\phi.
\laeq{entphaseret}
}
The retardation effect is manifest in this expression through the adiabatic functions $\Theta\left(t^{\rm sp}(\phi)\right)$ and $\Theta\left(t^{\rm ps}(\phi)\right)$, which restrict the entangling phase to the portion of the trajectory over which a gravitational signal can propagate between the source and the probe at the finite speed $c$. This causal structure is a genuine feature of the local, field-mediated description, kept explicit by the path integral formulation, and has no counterpart in Sec.~\ref{EntSch}, where the Lense-Thirring interaction enters the Hamiltonian~\eqref{eq:Ht} as an instantaneous function of the configuration. Under the locality and mediator assumptions standard in GIE, the detection of such retardation effects would provide stronger evidence for the existence of gravitons~\cite{mitrakos2026does}.

Now, if the source remains in the rotating superposition long enough that it remains in causal contact with the probe along the entire trajectory, that is, $\Theta\left(t^{\rm sp}(\phi)\right)=\Theta\left(t^{\rm ps}(\phi)\right)=1$, then the accumulated entangling phase reduces to
\alg{
\Delta=\frac{4\pi GJm_0}{\hbar Rc^2},
\laeq{DeltaPIEP}
}
which coincides with Eq.~\eqref{eq:entph2}, as expected.

\section{Conclusion}
\lsec{concl}

In this work, we have investigated gravity-induced entanglement generated by general relativistic frame dragging in the post-Newtonian regime through interferometry. In particular, we have evaluated the entanglement phase between the angular momentum of a rotating source mass prepared in a superposition and the arms of the interferometer by means of two complementary approaches. In the first approach, based on the Schr{\"o}dinger equation with a first-quantized Lense-Thirring Hamiltonian, we obtained the entanglement phase~\eqref{eq:entph2}, in agreement with the predictions of Refs.~\cite{lantano2024low,Petruzziello2025} in the semi-classical limit of a very large number of angular-momentum quanta for the source. In the second approach, based on the on-shell action of linearized quantum gravity coupled to matter, we recovered the same phase by performing a stationary-phase approximation over both the path of the probe particle and the gravitational field degrees of freedom. This result also coincides with the entanglement phase obtained by applying the general formula for the proper-time difference established in Ref.~\cite{wakakuwa2025detectability} to the present setup.

In this way, we have explicitly shown that the large-angular-momentum limit in the first-quantized description, \emph{i.e.}, the continuum limit, and the stationary-phase approximation in the path-integral formalism are mutually consistent, as both correspond to the same quasi-classical regime. At the same time, the path-integral derivation goes beyond the first-quantized description by clarifying the role played by retardation in the entanglement-generation process, thereby making consistency with relativistic causality manifest.

These findings provide the post-Newtonian counterpart of the result obtained by Christodoulou \emph{et al.}~\cite{christodoulou2023locally} in the Newtonian regime. Indeed, the same entangling phase derived from standard quantum-mechanical evolution generated by a Lense-Thirring potential is reproduced, within a manifestly local and relativistic framework, by the on-shell action of the linearized gravitational field. Hence, if the entanglement predicted in the present setup were detected, a plausible interpretation would be that it is mediated by linearized quantum gravity: a local and generally covariant description in which the gravitational field degrees of freedom are allowed to participate in the quantum dynamics.
Furthermore, the retardation effect in the entanglement-generation process is evident in our path-integral result \req{entphaseret}. As argued in \cite{mitrakos2026does}, observing such effects would provide stronger evidence for the existence of gravitons from the detection of GIE.

In the path-integral formulation, this interpretation becomes especially transparent. Since the stationary metric perturbation is sourced by the matter configuration in each branch, a source prepared in a superposition of angular momenta corresponds, within linearized quantum gravity, to branch-dependent weak-field geometries. In this sense, the present setup can be naturally interpreted as involving a superposition of frame dragging geometries. This goes beyond merely describing gravity as an effective non-classical channel, because the gravitational field degrees of freedom are explicitly retained before being integrated out. Importantly, however, such an observation should not be overstated: detecting the predicted entangling phase would probe the quantum nature of the gravitational interaction as a mediator, but it would not by itself amount to a direct observation of genuinely field-theoretic quantum aspects of gravity, such as propagating gravitons.

Finally, we stress that the significance of this work is not merely the derivation of the same entangling phase through two different methods, but the demonstration that the correspondence between effective gravitational entangling Hamiltonians and local, relativistically causal linearized-gravity descriptions extends beyond the Newtonian regime into the post-Newtonian gravitomagnetic sector. This opens the possibility of systematically studying relativistic corrections to gravity-induced entanglement
within a unified local-mediator framework.

A natural next step is to refine the analysis of the experimental conditions under which the above entanglement phase becomes accessible in the laboratory and to explore the role of relativistic retardation in suppressing or enhancing the generated correlations beyond the limiting cases discussed here. We leave a more detailed investigation of these aspects to future work.

\section*{Acknowledgement}
This work was supported by the ERC Synergy Grant HyperQ (Grant No. 856432), the DFG via QuantERA project LemaQume (Grant No. 500314265) and MEXT Quantum Leap Flagship Program (MEXT QLEAP), Grant No. JPMXS0120319794.  L.\ P.\ acknowledges networking support by the COST Actions CA23130 (Bridging high and low energies in search of quantum gravity), CA23115 (Relativistic Quantum Information) and CA24101 (Testing Fundamental Physics with Seismology).


\appendix

\begin{widetext}

\section{Calculation of the action integral}
\lapp{calaction}

\subsection{Calculation of $S^{sp}$}

Substituting the energy-momentum tensor of the probe particle, Eq.~\req{Tclock}, into the integral \req{actionintsc}, 
we have
\alg{
S^{\rm sp}=\frac{m_0}{4}\int dtd^3{\bm r} h_{\mu\nu}^{\rm source}({\bm r},t){V}_{\rm probe}^{\mu\nu}(t)\delta^{(3)}({\bm r}-{\bm q}(t)).
}
From the metric perturbation \req{hsource1}-\req{hsource3}, it follows that
\alg{
h_{\mu\nu}^{\rm source}(t,{\bm r}){V}_{\rm probe}^{\mu\nu}(t)
=
\frac{2GM\gamma(t)}{|{\bm r}|}\left(1+\frac{v(t)^2}{c^2}\right)
+
\frac{4GJ(t-|{\bm r}|/c)\gamma(t)}{c^2|{\bm r}|^3}(yv_x(t)-xv_y(t)).
}
Hence, we obtain $S^{\rm sp}=S_0^{\rm sp}+S_J^{\rm sp}$, where
\alg{
S_0^{\rm sp}=\frac{m_0}{4}\int dtd^3{\bm r} \frac{2GM\gamma(t)}{|{\bm r}|}\left(1+\frac{v(t)^2}{c^2}\right)\delta^{(3)}({\bm r}-{\bm q}(t))
}
and
\alg{
S_J^{\rm sp}=\frac{m_0}{4}\int dtd^3{\bm r} 
\frac{4GJ(t-|{\bm r}|/c)\gamma(t)}{c^2|{\bm r}|^3}(yv_x(t)-xv_y(t))\delta^{(3)}({\bm r}-{\bm q}(t)).
}
The effect of $S_0^{\rm sp}$ does not depend on $J$, and will be compensated in the two paths.
The only contribution to the phase shift is due to $S_J^{\rm sp}$.
It is calculated to be
\alg{
S_J^{\rm sp}
&=\frac{GJm_0}{c^2}\int 
\frac{\Theta(t({\bm q})-|{\bm q}(t)|/c)\gamma(t)}{|{\bm q}(t)|^3}(y_p(t)dx_p-x_p(t)dy_p) \nonumber\\
&=-\frac{GJm_0}{c^2}\int 
\Theta(t({\bm q})-|{\bm q}|/c)\gamma(t)\frac{({\bm q}\times d{\bm q})_z}{|{\bm q}|^3},
}
where $t({\bm q})$ denotes the inverse function of ${\bm q}(t)$, and the subscript $z$ denotes the $z$-component of the vector.
In the slow-motion approximation in which terms of order $v/c$ are neglected, we have $\gamma\approx1$ and thus
\alg{
S_J^{\rm sp}
&=
-\frac{GJm_0}{c^2}\int 
\Theta(t({\bm q})-|{\bm q}|/c)\frac{({\bm q}\times d{\bm q})_z}{|{\bm q}|^3}.
}

\subsection{Calculation of $S^{ps}$}

From \req{hclock} and \req{actionint}, we have 
\alg{
S^{\rm ps}=\frac{Gm_0}{c^4}\int dtd^3{\bm r} \frac{T_{\mu\nu}^{\rm source}(t,{\bm r})\bar{V}^{\mu\nu}_{\rm probe}(\tilde{t})}{|\tilde{{\bm d}}|-\tilde{{\bm d}}\cdot{\bm v}(\tilde{t})/c},
}
where $\tilde{t}=\tilde{t}(t,{\bm r})$ is the solution of $c(t-\tilde{t})=|{\bm r}-{\bm q}(\tilde{t})|$, and $\tilde{{\bm d}}=\tilde{{\bm d}}(t,{\bm r})={\bm r}-{\bm q}(\tilde{t}(t,{\bm r}))$.
Using \req{sourceT1} and \req{sourceT2}, the numerator in the integral is calculated to be
\alg{
T_{\mu\nu}^{\rm source}(t,{\bm r})\bar{V}_{\rm probe}^{\mu\nu}(\tilde{t})
=
\gamma(\tilde{t})\left[\frac{1}{2}c^2(c^2+v(\tilde{t})^2)M
+c^2J(t)(v_y(\tilde{t})\partial_x-v_x(\tilde{t})\partial_y)\right]\delta^{(3)}({\bm r}).
}
Therefore, we have $S^{\rm ps}=S_0^{\rm ps}+S_J^{\rm ps}$, where
\alg{
S_0^{\rm ps}=\frac{GMm_0}{2}\int dtd^3{\bm r} \frac{\gamma(\tilde{t})}{|\tilde{{\bm d}}|-\tilde{{\bm d}}\cdot{\bm v}(\tilde{t})/c}\left(1+\frac{v(\tilde{t})^2}{c^2}\right)\delta^{(3)}({\bm r})
\laeq{s0cs}
}
and
\alg{
S_J^{\rm ps}&=\frac{GJm_0}{c^2}\int dtd^3{\bm r} \frac{\gamma(\tilde{t})}{|\tilde{{\bm d}}|-\tilde{{\bm d}}\cdot{\bm v}(\tilde{t})/c}\cdot
\Theta(t)[v_y(\tilde{t})\partial_x-v_x(\tilde{t})\partial_y]\delta^{(3)}({\bm r}).
}
The latter can further be calculated to be
\alg{
S_J^{\rm ps}&=
\frac{GJm_0}{c^2}\int dtd^3{\bm r} \delta^{(3)}({\bm r})\Theta(t)
\left[\partial_y\left(\frac{v_x(\tilde{t})\gamma(\tilde{t})}{|\tilde{{\bm d}}|-\tilde{{\bm d}}\cdot{\bm v}(\tilde{t})/c}\right)-\partial_x\left(\frac{v_y(\tilde{t})\gamma(\tilde{t})}{|\tilde{{\bm d}}|-\tilde{{\bm d}}\cdot{\bm v}(\tilde{t})/c}\right)\right]\nonumber\\
&=
\frac{GJm_0}{c^2}\int dt\:\Theta(t)
\left.\left[\partial_y\left(\frac{v_x(\tilde{t})\gamma(\tilde{t})}{|\tilde{{\bm d}}|-\tilde{{\bm d}}\cdot{\bm v}(\tilde{t})/c}\right)-\partial_x\left(\frac{v_y(\tilde{t})\gamma(\tilde{t})}{|\tilde{{\bm d}}|-\tilde{{\bm d}}\cdot{\bm v}(\tilde{t})/c}\right)\right]\right|_{{\bm r}={\bm 0}}.
}
Let $\bar{t}(t)$ be the solution of $c(t-\bar{t})=|{\bm q}(\bar{t})|$.
As we prove below, it holds that
\alg{
&\frac{dt}{d\bar{t}}\left.\left[\partial_y\left(\frac{v_x(\tilde{t})\gamma(\tilde{t})}{|\tilde{{\bm d}}|-\tilde{{\bm d}}\cdot{\bm v}(\tilde{t})/c}\right)-\partial_x\left(\frac{v_y(\tilde{t})\gamma(\tilde{t})}{|\tilde{{\bm d}}|-\tilde{{\bm d}}\cdot{\bm v}(\tilde{t})/c}\right)\right]\right|_{{\bm r}={\bm 0}}
\nn\\
&=\frac{-\gamma(\bar{t})}{|{\bm q}(\bar{t})|+{\bm v}(\bar{t})\cdot{\bm q}(\bar{t})/c}\left(\frac{{\bm q}(\bar{t})}{|{\bm q}(\bar{t})|}\times\left\{\left[
\frac{\gamma(\bar{t})^2{\bm v}(\bar{t})\cdot{\bm a}(\bar{t})}{c^3}
+\frac{1+(v(\bar{t})/c)^2+{\bm q}(\bar{t})\cdot{\bm a}(\bar{t})/c^2}{|{\bm q}(\bar{t})|+{\bm v}(\bar{t})\cdot{\bm q}(\bar{t})/c}\right]{\bm v}(\bar{t})+\frac{1}{c}{\bm a}(\bar{t})\right\}\right)_{z}.
\laeq{pxpyvv}
}
In the slow-motion approximation in which $v/c$ is ignored, it follows that
\alg{
\frac{dt}{d\bar{t}}\left.\left[\partial_y\left(\frac{v_x(\tilde{t})\gamma(\tilde{t})}{|\tilde{{\bm d}}|-\tilde{{\bm d}}\cdot{\bm v}(\tilde{t})/c}\right)-\partial_x\left(\frac{v_y(\tilde{t})\gamma(\tilde{t})}{|\tilde{{\bm d}}|-\tilde{{\bm d}}\cdot{\bm v}(\tilde{t})/c}\right)\right]\right|_{{\bm r}={\bm 0}}
=-\left(\frac{{\bm q}(\bar{t})\times{\bm v}(\bar{t})}{|{\bm q}(\bar{t})|^3}\right)_z.
}
Thus, we have
\alg{
S_J^{\rm ps}
&=
-\frac{GJm_0}{c^2}\int d\bar{t}\:
\Theta(\bar{t}+|{\bm q}(\bar{t})|/c)\frac{({\bm q}(\bar{t})\times{\bm v}(\bar{t}))_z}{|{\bm q}(\bar{t})|^3}\nonumber\\
&=
-\frac{GJm_0}{c^2}\int 
\Theta(t({\bm q})+|{\bm q}|/c)\frac{({\bm q}\times d{\bm q})_z}{|{\bm q}|^3}.
}

\subsection{Proof of Eq.~\req{pxpyvv}}
\lapp{calcScs}

To prove \req{pxpyvv}, we start with the definition of $\tilde{t}$:
\alg{
c(t-\tilde{t})=|{\bm r}-{\bm q}(\tilde{t})|.
}
Taking the total differential of the above equation yields
\alg{
c(dt-d\tilde{t})=\frac{{\bm r}-{\bm q}(\tilde{t})}{|{\bm r}-{\bm q}(\tilde{t})|}\cdot(d{\bm r}-{\bm v}(\tilde{t})d\tilde{t}),
}
which gives
\alg{
\left(c-\frac{({\bm r}-{\bm q}(\tilde{t}))\cdot{\bm v}(\tilde{t})}{|{\bm r}-{\bm q}(\tilde{t})|}\right)d\tilde{t}
=
cdt-\frac{({\bm r}-{\bm q}(\tilde{t}))\cdot d{\bm r}}{|{\bm r}-{\bm q}(\tilde{t})|}.
\laeq{cdttil}
}
Hence,
\alg{
\frac{\partial\tilde{t}}{\partial x}
=
-\frac{x-x_p(\tilde{t})}{c|{\bm r}-{\bm q}(\tilde{t})|-({\bm r}-{\bm q}(\tilde{t}))\cdot{\bm v}(\tilde{t})}
=
-\frac{1}{c}\frac{x-x_p(\tilde{t})}{|\tilde{{\bm d}}|-\tilde{{\bm d}}\cdot{\bm v}(\tilde{t})/c}.
\laeq{ptiltpx}
}
We then have
\alg{
\frac{\partial v_i(\tilde{t})}{\partial x}
=
a_i(\tilde{t})\frac{\partial\tilde{t}}{\partial x}
}
and
\alg{
\frac{\partial \gamma(\tilde{t})}{\partial x}
=
\frac{\partial v_j(\tilde{t})}{\partial x}\frac{\partial\gamma}{\partial v_j}=\frac{\gamma(\tilde{t})^3v^j(\tilde{t})a_j(\tilde{t})}{c^2}\frac{\partial\tilde{t}}{\partial x}
=\frac{\gamma(\tilde{t})^3{\bm v}(\tilde{t})\cdot{\bm a}(\tilde{t})}{c^2}\frac{\partial\tilde{t}}{\partial x},
}
which yields
\alg{
\frac{\partial}{\partial x}\left(v_y(\tilde{t})\gamma(\tilde{t})\right)
=
\left[
a_y(\tilde{t})\gamma(\tilde{t})+\frac{\gamma(\tilde{t})^3{\bm v}(\tilde{t})\cdot{\bm a}(\tilde{t})}{c^2}v_y(\tilde{t})
\right]
\frac{\partial\tilde{t}}{\partial x}.
\laeq{pvygpx}
}
Additionally, we have
\alg{
\frac{\partial \tilde{d}_i}{\partial x}
=
\delta_{xi}-v_i\frac{\partial\tilde{t}}{\partial x},
}
\alg{
\frac{\partial |\tilde{d}|}{\partial x}
=
\frac{\tilde{d}^j}{|\tilde{d}|}\frac{\partial \tilde{d}^j}{\partial x}
=
\frac{\tilde{d}^j}{|\tilde{d}|}\left(\delta_{xj}-v_j\frac{\partial\tilde{t}}{\partial x}\right)
=
\frac{\tilde{d}_x}{|\tilde{d}|}
-
\frac{\tilde{{\bm d}}\cdot{\bm v}(\tilde{t})}{|\tilde{d}|}\frac{\partial\tilde{t}}{\partial x}
}
and
\alg{
\frac{\partial(\tilde{{\bm d}}\cdot{\bm v}(\tilde{t}))}{\partial x}
=
\frac{\partial\tilde{d}_j}{\partial x}v^j+\tilde{d}_j\frac{\partial v^j}{\partial x}
=v_x-v^2\frac{\partial\tilde{t}}{\partial x}+\tilde{{\bm d}}\cdot{\bm a}(\tilde{t})\frac{\partial\tilde{t}}{\partial x},
}
which implies
\alg{
\frac{\partial}{\partial x}\left(|\tilde{{\bm d}}|-\frac{\tilde{{\bm d}}\cdot{\bm v}(\tilde{t})}{c}\right)
=
\frac{\tilde{d}_x}{|\tilde{d}|}-\frac{v_x}{c}-\left(\frac{\tilde{{\bm d}}\cdot{\bm v}(\tilde{t})}{|\tilde{d}|}+\frac{v^2}{c}-\frac{\tilde{{\bm d}}\cdot{\bm a}(\tilde{t})}{c}\right)\frac{\partial\tilde{t}}{\partial x}.
\laeq{pddvcpx}
}

Now we substitute $\tilde{t}|_{{\bm r}={\bm 0}}=\bar{t}$ and $\tilde{{\bm d}}|_{{\bm r}={\bm 0}}=-{\bm q}(\bar{t})$ into \req{ptiltpx}, \req{pvygpx} and \req{pddvcpx}.
Noting that
\alg{
\left.\left(|\tilde{{\bm d}}|-\frac{\tilde{{\bm d}}\cdot{\bm v}(\tilde{t})}{c}\right)\right|_{{\bm r}={\bm 0}}
=
\frac{1}{c}\left(c|{\bm q}(\bar{t})|+{\bm q}(\bar{t})\cdot{\bm v}(\bar{t})\right),
}
we have
\alg{
\left.\frac{\partial\tilde{t}}{\partial x}\right|_{{\bm r}={\bm 0}}
&=
\frac{x_p(\bar{t})}{c|{\bm q}(\bar{t})|+{\bm v}(\bar{t})\cdot{\bm q}(\bar{t})},\\
\left.\frac{\partial}{\partial x}\left(v_y(\tilde{t})\gamma(\tilde{t})\right)\right|_{{\bm r}={\bm 0}}
&=
\frac{x_p(\bar{t})}{c|{\bm q}(\bar{t})|+{\bm v}(\bar{t})\cdot{\bm q}(\bar{t})}\left[
a_y(\bar{t})\gamma(\bar{t})+\frac{\gamma(\bar{t})^3{\bm v}(\bar{t})\cdot{\bm a}(\bar{t})}{c^2}v_y(\bar{t})
\right]
}
and
\alg{
&\left.\frac{\partial}{\partial x}\left(|\tilde{{\bm d}}|-\frac{\tilde{{\bm d}}\cdot{\bm v}(\tilde{t})}{c}\right)\right|_{{\bm r}={\bm 0}}
=
-\frac{x_p(\bar{t})}{|{\bm q}(\bar{t})|}-\frac{v_x(\bar{t})}{c}+\left(\frac{{\bm q}(\bar{t})\cdot{\bm v}(\bar{t})}{|{\bm q}(\bar{t})|}-\frac{v^2}{c}-\frac{{\bm q}(\bar{t})\cdot{\bm a}(\bar{t})}{c}\right)\frac{x_p(\bar{t})}{c|{\bm q}(\bar{t})|+{\bm v}(\bar{t})\cdot{\bm q}(\bar{t})}.
}
Combining these together, we obtain
\alg{
&\left.\left[\left(|\tilde{{\bm d}}|-\frac{\tilde{{\bm d}}\cdot{\bm v}(\tilde{t})}{c}\right)\frac{\partial}{\partial x}\left(v_y(\tilde{t})\gamma(\tilde{t})\right)
-
\left(v_y(\tilde{t})\gamma(\tilde{t})\right)\frac{\partial}{\partial x}\left(|\tilde{{\bm d}}|-\frac{\tilde{{\bm d}}\cdot{\bm v}(\tilde{t})}{c}\right)\right]\right|_{{\bm r}={\bm 0}}
\nn\\
&
=
\frac{1}{c}\left[
a_y(\bar{t})\gamma(\bar{t})+\frac{\gamma(\bar{t})^3{\bm v}(\bar{t})\cdot{\bm a}(\bar{t})}{c^2}v_y(\bar{t})
\right]
x_p(\bar{t})\nn\\\nonumber
&\quad
+
\left(v_y(\bar{t})\gamma(\bar{t})\right)\left[\frac{x_p(\bar{t})}{|{\bm q}(\bar{t})|}+\frac{v_x(\bar{t})}{c}+\left(-\frac{{\bm q}(\bar{t})\cdot{\bm v}(\bar{t})}{|{\bm q}(\bar{t})|}+\frac{v(\bar{t})^2}{c}+\frac{{\bm q}(\bar{t})\cdot{\bm a}(\bar{t})}{c}\right)\frac{x_p(\bar{t})}{c|{\bm q}(\bar{t})|+{\bm v}(\bar{t})\cdot{\bm q}(\bar{t})}\right]\\
&=\left\{\frac{1}{c}\left[
a_y(\bar{t})\gamma(\bar{t})+\frac{\gamma(\bar{t})^3{\bm v}(\bar{t})\cdot{\bm a}(\bar{t})}{c^2}v_y(\bar{t})
\right]+\frac{v_y(\bar{t})\gamma(\bar{t})}{|{\bm q}(\bar{t})|}\right.\nn\\\nonumber
&\quad\quad\left.+\left(-\frac{{\bm q}(\bar{t})\cdot{\bm v}(\bar{t})}{|{\bm q}(\bar{t})|}+\frac{v(\bar{t})^2}{c}+\frac{{\bm q}(\bar{t})\cdot{\bm a}(\bar{t})}{c}\right)\frac{v_y(\bar{t})\gamma(\bar{t})}{c|{\bm q}(\bar{t})|+{\bm v}(\bar{t})\cdot{\bm q}(\bar{t})}\right\}x_p(\bar{t})+\frac{\gamma(\bar{t})v_x(\bar{t})v_y(\bar{t})}{c}
\\
&=\left\{
\frac{\gamma(\bar{t})^3{\bm v}(\bar{t})\cdot{\bm a}(\bar{t})}{c^3}
+\frac{\gamma(\bar{t})}{|{\bm q}(\bar{t})|}+\left(-\frac{{\bm q}(\bar{t})\cdot{\bm v}(\bar{t})}{|{\bm q}(\bar{t})|}+\frac{v(\bar{t})^2}{c}+\frac{{\bm q}(\bar{t})\cdot{\bm a}(\bar{t})}{c}\right)\frac{\gamma(\bar{t})}{c|{\bm q}(\bar{t})|+{\bm v}(\bar{t})\cdot{\bm q}(\bar{t})}\right\}x_p(\bar{t})v_y(\bar{t})\nn\\\nonumber
&\quad\quad+\frac{\gamma(\bar{t})a_y(\bar{t})x_p(\bar{t})}{c}+\frac{\gamma(\bar{t})v_x(\bar{t})v_y(\bar{t})}{c}\\
&=\left[
\frac{\gamma(\bar{t})^2{\bm v}(\bar{t})\cdot{\bm a}(\bar{t})}{c^2}
+\frac{c}{|{\bm q}(\bar{t})|}+\left(-\frac{c{\bm q}(\bar{t})\cdot{\bm v}(\bar{t})}{|{\bm q}(\bar{t})|}+v(\bar{t})^2+{\bm q}(\bar{t})\cdot{\bm a}(\bar{t})\right)\frac{1}{c|{\bm q}(\bar{t})|+{\bm v}(\bar{t})\cdot{\bm q}(\bar{t})}\right]\frac{\gamma(\bar{t})x_p(\bar{t})v_y(\bar{t})}{c}\nn\\\nonumber
&\quad\quad+\frac{\gamma(\bar{t})a_y(\bar{t})x_p(\bar{t})}{c}+\frac{\gamma(\bar{t})v_x(\bar{t})v_y(\bar{t})}{c}\\
&=\left[
\frac{\gamma(\bar{t})^2{\bm v}(\bar{t})\cdot{\bm a}(\bar{t})}{c^2}
+\frac{c^2+v(\bar{t})^2+{\bm q}(\bar{t})\cdot{\bm a}(\bar{t})}{c|{\bm q}(\bar{t})|+{\bm v}(\bar{t})\cdot{\bm q}(\bar{t})}\right]\frac{\gamma(\bar{t})x_p(\bar{t})v_y(\bar{t})}{c}+\frac{\gamma(\bar{t})x_p(\bar{t})a_y(\bar{t})}{c}+\frac{\gamma(\bar{t})v_x(\bar{t})v_y(\bar{t})}{c}.
}
Therefore, we arrive at
\alg{
&\left.\left[\partial_x\left(\frac{v_y(\tilde{t})\gamma(\tilde{t})}{|\tilde{{\bm d}}|-\tilde{{\bm d}}\cdot{\bm v}(\tilde{t})/c}\right)-\partial_y\left(\frac{v_x(\tilde{t})\gamma(\tilde{t})}{|\tilde{{\bm d}}|-\tilde{{\bm d}}\cdot{\bm v}(\tilde{t})/c}\right)\right]\right|_{{\bm r}={\bm 0}}
\nn\\
&=\frac{c\gamma(\bar{t})}{(c|{\bm q}(\bar{t})|+{\bm v}(\bar{t})\cdot{\bm q}(\bar{t}))^2}\left({\bm q}(\bar{t})\times\left\{\left[
\frac{\gamma(\bar{t})^2{\bm v}(\bar{t})\cdot{\bm a}(\bar{t})}{c^2}
+\frac{c^2+v(\bar{t})^2+{\bm q}(\bar{t})\cdot{\bm a}(\bar{t})}{c|{\bm q}(\bar{t})|+{\bm v}(\bar{t})\cdot{\bm q}(\bar{t})}\right]{\bm v}(\bar{t})+{\bm a}(\bar{t})\right\}\right)_z.
}
From \req{cdttil}, we have
\alg{
\left(c+\frac{{\bm q}(\bar{t})\cdot{\bm v}(\bar{t})}{|{\bm q}(\bar{t})|}\right)d\bar{t}
=
cdt,
}
which implies
\alg{
\frac{dt}{d\bar{t}}=1+\frac{{\bm v}(\bar{t})\cdot{\bm q}(\bar{t})}{c|{\bm q}(\bar{t})|}.
}
Thus
\alg{
&\left.\left[\partial_x\left(\frac{v_y(\tilde{t})\gamma(\tilde{t})}{|\tilde{{\bm d}}|-\tilde{{\bm d}}\cdot{\bm v}(\tilde{t})/c}\right)-\partial_y\left(\frac{v_x(\tilde{t})\gamma(\tilde{t})}{|\tilde{{\bm d}}|-\tilde{{\bm d}}\cdot{\bm v}(\tilde{t})/c}\right)\right]\right|_{{\bm r}={\bm 0}}\frac{dt}{d\bar{t}}\nonumber
\\
&=\frac{\gamma(\bar{t})}{c|{\bm q}(\bar{t})|+{\bm v}(\bar{t})\cdot{\bm q}(\bar{t})}\left(\frac{{\bm q}(\bar{t})}{|{\bm q}(\bar{t})|}\times\left\{\left[
\frac{\gamma(\bar{t})^2{\bm v}(\bar{t})\cdot{\bm a}(\bar{t})}{c^2}
+\frac{c^2+v(\bar{t})^2+{\bm q}(\bar{t})\cdot{\bm a}(\bar{t})}{c|{\bm q}(\bar{t})|+{\bm v}(\bar{t})\cdot{\bm q}(\bar{t})}\right]{\bm v}(\bar{t})+{\bm a}(\bar{t})\right\}\right)_z\nonumber
\\
&=\frac{\gamma(\bar{t})}{|{\bm q}(\bar{t})|+{\bm v}(\bar{t})\cdot{\bm q}(\bar{t})/c}\left(\frac{{\bm q}(\bar{t})}{|{\bm q}(\bar{t})|}\times\left\{\left[
\frac{\gamma(\bar{t})^2{\bm v}(\bar{t})\cdot{\bm a}(\bar{t})}{c^3}
+\frac{1+(v(\bar{t})/c)^2+{\bm q}(\bar{t})\cdot{\bm a}(\bar{t})/c^2}{|{\bm q}(\bar{t})|+{\bm v}(\bar{t})\cdot{\bm q}(\bar{t})/c}\right]{\bm v}(\bar{t})+\frac{1}{c}{\bm a}(\bar{t})\right\}\right)_z.
\laeq{prfLE}
}

\end{widetext}

\bibliography{bibliography.bib}

\end{document}

%% file: preferences.tex
{\theorembodyfont{\normalfont\it}
\theoremheaderfont{\normalfont\bf}
\newtheorem{thm}{ Theorem}
\newtheorem{dfn}[thm]{ Definition}
\newtheorem{lmm}[thm]{ Lemma}

\newtheorem{crl}[thm]{ Corollary}
\newtheorem{asm}[thm]{ Assumption}
\newtheorem{prp}[thm]{ Proposition}
\newtheorem{cjt}[thm]{ Conjecture}
\newtheorem{rmk}[thm]{ Remark}}

{\theorembodyfont{\normalfont}
\theoremheaderfont{\normalfont\it}
\newtheorem{prf}{ Proof:}}

{\theorembodyfont{\normalfont}
\theoremheaderfont{\normalfont\it}
}

{\theorembodyfont{\normalfont}
\theoremheaderfont{\normalfont\it}
}

{\theorembodyfont{\normalfont}
\theoremheaderfont{\normalfont\it}
}

{\theorembodyfont{\normalfont}
\theoremheaderfont{\normalfont\it}
}

{\theorembodyfont{\normalfont}
\theoremheaderfont{\normalfont\it}
}

\newcommand{\bra}[1]{\mbox{$\langle#1|$}}

\newcommand{\ket}[1]{\mbox{$|#1\rangle$}}

\newcommand{\proj}[1]{\mbox{$\ket{#1}\!\bra{#1}$}}

\newcommand{\alg}[1]{\begin{align}#1\end{align}}

\newcommand{\nn}{\nonumber}

\newcommand{\bthm}[1]{\begin{thm}\label{thm:#1}}
\newcommand{\ethm}{\end{thm}}

\newcommand{\rThm}[1]{Theorem \ref{thm:#1}}
\newcommand{\blmm}[1]{\begin{lmm}\label{lmm:#1}}
\newcommand{\elmm}{\end{lmm}}

\newcommand{\bdfn}[1]{\begin{dfn}\label{dfn:#1}}
\newcommand{\edfn}{\end{dfn}}

\newcommand{\basm}[1]{\begin{asm}\label{asm:#1}}
\newcommand{\easm}{\end{asm}}

\newcommand{\bprp}[1]{\begin{prp}\label{prp:#1}}
\newcommand{\eprp}{\end{prp}}

\newcommand{\bcrl}[1]{\begin{crl}\label{crl:#1}}
\newcommand{\ecrl}{\end{crl}}

\newcommand{\bcjt}[1]{\begin{cjt}\label{cjt:#1}}
\newcommand{\ecjt}{\end{cjt}}

\newcommand{\brmk}[1]{\begin{rmk}\label{rmk:#1}}
\newcommand{\ermk}{\end{rmk}}

\newcommand{\bprf}{\begin{prf}}
\newcommand{\eprf}{\end{prf}}
\newcommand{\laeq}[1]{\label{eq:#1}}
\newcommand{\req}[1]{(\ref{eq:#1})}

\newcommand{\lsec}[1]{\label{sec:#1}}

\newcommand{\lapp}[1]{\label{app:#1}}

\newcommand{\rApp}[1]{Appendix \ref{app:#1}}

\newcommand{\bitem}{\begin{itemize}}
\newcommand{\entem}{\end{itemize}}
\newcommand{\benum}{\begin{enumerate}}
\newcommand{\ennum}{\end{enumerate}}

\newcommand{\otm}{\otimes}
